\documentclass[letterpaper,twocolumn,10pt,hyphens]{article}
\usepackage{usenix2023_SOUPS}
\usepackage{epsfig,endnotes}
\usepackage{color,soul}
\usepackage{xcolor}
\newcommand{\etal}[0]{et~al{.}}

\begin{document}

\date{}

\title{\Large \bf Towards Equitable Privacy}

\def\plainauthor{Author name(s) for PDF metadata. Don't forget to anonymize for submission!}

\author{
{\rm Kopo M. Ramokapane}\\
University of Bristol
\and
{\rm Lizzie Coles-Kemp}\\
Royal Holloway University of London
\and
{\rm Nikhil Patnaik }\\
University of Bristol
\and
{\rm Rui Huan }\\
University of Bristol
\and
{\rm Nirav Ajmeri}\\
University of Bristol
\and
{\rm Genevieve Liveley}\\
University of Bristol
\and
{\rm Awais Rashid}\\
University of Bristol
}

\maketitle

\thecopyright

\begin{abstract}
Ensuring equitable privacy experiences remains a challenge, especially for marginalised and vulnerable populations (MVPs) who often hesitate to participate or use digital services due to concerns about the privacy of their sensitive information. In response, security research has emphasised the importance of inclusive security and privacy practices to facilitate meaningful engagement of MVPs online. However, research in this area is still in its early stages, with other MVPs yet to be considered (such as low-income groups, and refugees), novel engagement methods yet to be explored, and limited support for software developers in building applications and services for MVPs. In 2022, we initiated a UK Research Council funded Equitable Privacy project to address these gaps. Our goal is to prioritise the privacy needs and requirements of MVPs in the design and development of software applications and services. 
We design and implement a new participatory research approach -- community studybeds -- in collaboration with third-sector organisations that support MVPs to identify and tackle the challenges these groups encounter. In this paper, we share the initial reflections and experiences of the Equitable Privacy project, particularly emphasising the utilisation of our community studybeds.
\end{abstract}

\section{Introduction}
\label{sec:introduction}

While the right to privacy is often regarded as a universal entitlement, achieving equitable implementation of privacy principles online remains a significant challenge. Marginalised and vulnerable populations (MVPs) often abstain from online participation or using digital services due to fears surrounding the potential exposure of their sensitive information~\cite{mcdonald2021citizens}. Consequently, a growing body of security research has highlighted the importance of developing inclusive security and privacy practices to facilitate the meaningful engagement of MVPs online. For instance, Wang~\cite{wang2017third} calls for research work that empowers people with ``various characteristics, abilities, needs, and values,'' while Das Chowdhury~\etal~\cite{chowdhury2022utility} underscore the necessity of embracing and responding to these diversities when developing PETs using the capability approach. They also argue that the current way of assessing or designing PETs is more utility-based (i.e., focused on technical and usability aspects) and does not consider the realities of MVPs. Sannon and Forte~\cite{sannon2022privacy} further highlight that MVPs can have unique privacy needs and tend to experience disproportionate harm when their privacy is violated.

Consequently, a body of work has attempted to understand the security and privacy needs of MVPs. For instance, prior works~\cite{ahmed2015privacy,hayes2019cooperative,akter2020uncomfortable,napoli2021m} have explored privacy concerns and behaviours of people with visual impairments, emphasising that these issues arise because they have not been included in the design process. Others have focused on issues such as intimate partner violence (IPV), examining how technology is used to abuse~\cite{southworth2007intimate,fraser2010new,murray2015domestic,tseng2020tools} and how survivors protect themselves~\cite{arief2014sensible,freed2017digital,matthews2017stories}. Some works~\cite{Dantec2011Public,havron2019clinical,kuo2023understanding} have addressed the need to support service providers. Others~\cite{zhang2021webally,renaud2022accessible} have advocated for integrating accessibility into security and privacy tools. Despite these efforts, most technology continues to prioritise the needs of the masses, often overlooking or unintentionally excluding MVPs from their use cases. However, designing for the groups at the edges can also create solutions that benefit the broader population. It is essential to recognise that the definition of an MVP is nuanced; an individual may not be socio-economically disadvantaged but can still be a victim of IPV or surveillance, or suddenly become a refugee (as seen in recent events in Ukraine~\cite{UkraineRefugeeSituation} and Sudan~\cite{SouthSudanCrisisWatch}).




In our pursuit of narrowing the disparity between the general population and MVP in their access to privacy protections, last year, in 2022, we commenced the Equitable Privacy project. The Equitable Privacy project~\footnote{\url{https://gow.epsrc.ukri.org/NGBOViewGrant.aspx?GrantRef=EP/W025361/1}} aims to prioritise the privacy needs of marginalised and vulnerable populations in designing and developing software applications and services, thus bringing MVPs and developers together.

This paper presents our initial reflections and experiences striving for equitable privacy, employing community studybeds as our core participation research methodology. First, we discuss Equitable Privacy, focusing on inclusive technology and addressing the security and privacy risks MVPs face. Next, we share our experiences setting up community studybeds, highlighting the methodology and insights gained from the process.

\section{Equitable Privacy (EP)}
\label{sec:equitable-privacy}
\emph{Equitable privacy} is a conceptual framework that aims to ensure the just and fair provision of privacy to all individuals, regardless of their social, economic, and demographic backgrounds. The framework recognises that privacy experiences are not uniform and that certain individuals or communities are more vulnerable or disadvantaged, facing unique challenges or vulnerabilities regarding privacy protection. For instance, individuals in abusive and coercive relationships, refugees, or political activists have nuanced privacy and information control needs~\cite{parkin2019usability, albrecht2021collective}. While a monitoring app may be supportive in certain settings, such as healthcare, it can become a means of oppression for these user groups~\cite{freed2018stalker,rahul2018spyware}. EP also recognises that the design of privacy mechanisms, lack of transparency, accessibility or accountability in how data is utilised, often leads to distrust and disenfranchisement. This can create a perception of privacy mechanisms being turned against them\textemdash victims of sexual assault, for example, have expressed a lack of trust in online reporting systems due to fears about privacy, anonymity, and traceability~\cite{obieh2020towards}. 

Another example pertains to individuals and groups that experience barriers to access~\cite{GovUKBarriers}. There is a growing recognition that many individuals and groups face barriers to digital access such as financial constraints, limited accessibility, capacity limitations, and socio-cultural factors~\cite{Carpenter2019,Schauberger2023}. Consequently, security practitioner communities are increasingly considering how these barriers affect how individuals and groups access and protect information~\cite{NCSC2021c, coles2022protecting}, as these barriers can have significant implications for informational privacy. For example, the barriers to access may result in individuals sharing devices to access essential services involving sensitive personal information, such as healthcare, welfare, finance, and victim support. Alternatively, they may rely on assistance from friends and family~\cite{coles2022protecting}. While such support is often beneficial, it can also result in fraud and harms~\cite{kemp2023consumer}. These issues not only heighten insecurity for individuals already experiencing socioeconomic, emotional, physical, or political precarity, but they may also impede digital participation or the adoption of digital and privacy technologies.

The notion of equitable privacy recognises that not only certain dimensions of identity, such as race, ability, ethnicity, gender, age, and socio-economic status, often introduce disparities and inequalities in privacy protections but also the intersections between these dimensions can simultaneously both amplify and hide these disparities and inequalities. The EP framework highlights the pressing need to identify and mitigate the privacy-related risks and harms that may disproportionately affect marginalised or disadvantaged groups.

\section{Community Studybeds}
\label{sec:studybeds}
\vspace{-6pt}

To better understand the privacy needs and challenges of MVPs, we engage with them through Community studybeds. Such studybeds serve as sites of co-investigation and exploration, building upon established frameworks such as Living Labs and Testbeds. These frameworks involve multiple stakeholders and focus on co-creating innovation in real-world contexts~\cite{LivingLabs, Cordis}. However, community studybeds differentiate themselves by utilising a participatory design approach~\cite{vines2013configuring} that places people and their privacy concerns at the core of the study design. The study contexts are established in consultation with the participant groups, ensuring relevance and alignment with their experiences. Moreover, a community researchbed approach emphasises establishing partnerships with community groups, including third-sector organisations, with a shared emphasis on capacity building. Rather than treating community groups as passive participants, they are considered active partners co-designing research direction as well as actively participating in the research. The timing and pace of the community researchbed activities are also determined in collaboration with the participant groups, allowing for a more inclusive and participatory approach~\cite{coles2013letting}. We have currently established three community studybeds with four different organisations in two locations: one organisation in Sunderland and three organisations in Bristol.

\subsection{Sunderland Community Group}
At the time of writing, one community researchbed had been established in Sunderland, North East England with a voluntary organisation that takes the role of research partner. The inquiry focus of this researchbed is digitally-enabled scams, and an initial engagement has been completed using the Neighbourhood Ideas Exchange toolkit from public goods lab, Proboscis. This consultation enabled us to discuss how digitally-enabled scams appear in day-to-day life, their impact on participants' daily lives, and the resulting adverse consequences. The participants included representatives from four voluntary and third sector and local government organisations. The principle of equity is core to both the community researchbed design and to the processes of establishing and carrying out the equitable privacy inquiry.

\textbf{Equity in focus and context:} During the initialisation of the community researchbed, researchers worked with community workers and representatives from participant groups to establish the relevant context for an equitable privacy inquiry. It was agreed that digitally-driven fraud and scams was the most appropriate context because they represent a constant pressure that affects everyday digital interactions. 

\textbf{Equity in design and process:} Following participatory design principles, the participant groups shaped the subsequent engagements for the inquiry, set out the reciprocity arrangement (i.e., the benefits that the individuals and groups would receive in return for taking part), and the timings of the engagements.

\textbf{Equity in outputs and dissemination:} As part of the reciprocity agreement, the participant groups and the research partner organisation takes an active role in the research analysis and in the dissemination process for the outputs. The community researchbed inquiry will next move to a wider community engagement. The host organisation has designed a community information package and between July and August 2023 will lead scams and fraud awareness and discussion sessions.  The data analysis will be co-developed with the research partner and participant groups and be used to shape equitable privacy interventions.

\subsection{Bristol Community Groups}
We have established two community studybeds hosted by three voluntary organisations that work with different communities in Bristol, South West England. The first commmunity researchbed in Bristol focuses on energy and the associated risks related to energy management systems. It is hosted by two voluntary organisations. Organisation~A utilises technology and the arts to generate creative solutions, ensuring the inclusion of individuals and groups at risk of social and digital exclusion. Organisation~B tackles energy issues in Bristol by engaging individuals and community groups with an interest in energy. The second community researchbed is hosted by an organisation (Organisation~C) that is specifically dedicated to working with survivors of sexual abuse.

Regarding the first community researchbed, our initial engagement with the partner organisations began with meetings to understand the services they offer and the community they serve. During this time, we also shared the goals of our project and what we hope to achieve. In our second meeting with Organisation~A, we introduced the community workers to a tabletop game called ``Decisions and Disruptions~\footnote{https://www.decisions-disruptions.org/}.'' This game, developed by our research group, challenges players to manage the security of a small utility company with a given budget. The game presents various security scenarios, requiring players to consider potential threats, infrastructure vulnerabilities, past and ongoing cyber-attacks, and budget limitations. This activity not only helped build rapport and highlight our potential contribution to the partnership but also raised security awareness among the community workers~\cite{frey2017good,shreeve2020so}. Regarding the second community researchbed, we have only met with partner Organisation~C. This engagement established the context of our inquiry and discussed the conduct of research engagements and the responsibilities of each partner.

Similar to our work in Sunderland, our goal in the initial engagements was to ensure fairness and equal opportunities for our partner organisations in establishing the community research bed and investigating the issues at hand.

\textbf{Equity in focus and context:} Since both Organisation~A and Organisation~B were already involved in energy projects at various capacities, the researchers met and discussed their respective projects to identify common interests and potential benefits for both parties. With Organisation~A, the researchers and community workers explored how community members could be encouraged to share their energy-related data through a community dashboard. On the other hand, the researchers and Organisation~B agreed to organise energy awareness clinics, during which the researchers would focus on understanding the community members' concerns regarding energy-related technologies while the community workers would raise awareness about effective energy management. 

Our initial engagement with Organisation~C followed a similar pattern. The researchers shared information about their ongoing projects on online citizen protection while the community workers described their work with survivors of sexual abuse. Both parties agreed to focus on issues concerning the sharing of digital material as evidence after reporting abuse.

\textbf{Equity in design and process:}
In collaboration with Organisation~A, the researchers organised the first workshop on developing the community dashboard. The community workers took the lead in planning, deciding on the inquiry method, recruitment process, and workshop date. Since Organisation~B was already conducting workshops with various groups in Bristol, the community workers shared their event calendar with the researchers, and together they identified which workshops would be utilised as energy clinics for the studies. In the initial meeting with Organisation~C, the community workers shared ideas with the researchers on how both parties could collaborate for mutual benefit. Discussions included engagement methods with community members, the duration of these engagements, and the scheduling of activities.

\textbf{Equity in outputs and dissemination:}
Following the initial workshop, Organisation~A collected and took the lead in analysing the workshop materials. The community workers analysed the data and prepared an online board to share the key outputs of the workshop. Prior to releasing the findings, both parties held a debrief meeting to reflect on the workshop and discuss the findings.

\subsection{Developer Panel}
As part of our Equitable Privacy project, we aim to support developers in designing and developing software applications and services that enable equitable privacy experiences. To achieve this, we are currently working on establishing a developer panel to identify and address technological gaps in developing applications and services for MVPs. 


We are currently in the process of assembling a panel by leveraging our connections with industry professionals and software development communities that we have established through our previous projects. Also, we will invite developers who voluntarily engage with MVPs in their own time to join the panel. This diverse panel, comprising developers with varied project experiences and a range of end-users for whom they have developed applications, will offer unique perspectives on privacy, fairness, and the specific needs of MVPs. It will also open up new avenues for research. The panel will also shed light on the challenges developers face as we study them using API features and existing privacy tools. Similar to our approach to the community studybeds, we intend to ensure equity in the context, design of activities, and dissemination of outputs through close collaboration with the developer panel.



\section{Initial Lessons from Establishing Community Studybeds}
\label{sec:lessons}

\textbf{Partnerships.}
Enabling equitable privacy experiences requires partnerships between research partners, community workers, and the groups they serve. In setting up community studybeds, engaging community representatives as partners has provided us with a deeper understanding of the issues they address in the community, the existing disparities, and how we can effectively engage different participation groups. It has also helped us contextualise the focus of our studies, design our inquiries to align with the practical needs of running activities with community groups, and makes the process of engagement more accessible for participants. 

\textbf{A deeper understanding of vulnerability is necessary.} Researchers often approach studies and issues related to MVPs with their understanding of who is considered vulnerable. However, working with our partner organisations has highlighted that while there are commonalities in the concept of ``vulnerability'' across various groups and organisations, it can have subtle differences in meaning. For example, Organisation B defined \textit{vulnerability} as anyone struggling to pay their energy bills, whereas Organisation A may have a different perspective. It is crucial for researchers to avoid imposing their definitions and instead work closely with community workers to understand the meaning within each specific context.

\textbf{Considerations for interviews.} In typical privacy studies, conducting interviews with participants is often seen as a routine practice without significant concerns. However, our partner organisations have emphasised the importance of considering the comfort levels of community groups during interviews. For instance, participants may feel uncomfortable sharing their experiences with a researcher who resembles their abuser (e.g., a male interviewer interviewing a woman). By working in partnership with organisations, we can identify these nuanced issues that may not be apparent if community workers and groups are merely treated as participants.

\textbf{More than just study activities.} We have also learned that to enhance engagement from community groups, it is essential to consider the needs of individuals whose participation may be influenced by the presence of others accompanying them. For example, organising workshops may require arrangements for childminders or providing engaging activities for accompanying individuals. Recognising that some people may have other responsibilities that prevent their participation is crucial in fostering inclusivity, and understanding the diverse circumstances of community members.

\section{Limitations}
\label{sec:limitations}

An equitable approach does not necessarily result in an equitable outcome. The power imbalances between users of technology and the technology companies are not swept away by this approach. Furthermore, the principles of an equitable are often challenging to fully implement. Whilst the principles of voluntary participation, reciprocity, and context design and selection are intended to be in the hands of research partners and the community resesarchbed participants, the social dynamics of the community researchbed mean that these ideals are not always fully realised. However, such an approach does offer a step towards making user-centred privacy research fairer and more just.

\section{Conclusion}
\label{sec:conclusion}
We presented our initial reflections and experiences of the Equitable Privacy project, focusing on using community studybeds as a participatory research methodology. Taking this approach, the community researchbed becomes a space in which individuals can voice concerns regarding equity, influence the direction of the inquiry, and guide the selection of interventions. The use of community studybeds highlights the effectiveness of partnership in understanding the privacy needs of MVPs for designing and developing software and services that prioritise equitable privacy experiences.

\section*{Acknowledgments}
This work is generously funded the EPSRC (EP/W025361/1).

\bibliographystyle{plain}
\bibliography{References}

\end{document}